\documentclass[12pt]{iopart}

\usepackage{graphicx}
\usepackage{epstopdf}
\usepackage{enumerate}
\usepackage{subfig}

\newcommand{\diff}[1]{\mathrm{d}{#1}}
\newcommand{\de}[1]{\delta #1}

\newcommand{\pder}[2]{\frac{\partial{#1}}{\partial{#2}}}
\newcommand{\pdermin}[2]{{\partial}_{#2}{#1}}

\begin{document}

\title[Electrogeodesics in the di-hole Majumdar-Papapetrou spacetime]{Electrogeodesics in the di-hole Majumdar-Papapetrou spacetime}

\author{Ji\v{r}\'{\i} Ryzner \& Martin \v{Z}ofka}

\address{Institute of Theoretical Physics, Charles University in Prague}
\ead{j8.ryzner@gmail.com, zofka@mbox.troja.mff.cuni.cz}
\vspace{10pt}

\begin{abstract}
We investigate the (electro-)geodesic structure of the Majumdar-Papapetrou solution representing static charged black holes in equilibrium. We assume only two point sources, imparting thus the spacetime axial symmetry. We study electrogeodesics both in and off the equatorial plane and explore the stability of circular trajectories via geodesic deviation equation. In contrast to the classical Newtonian situation, we find regions of spacetime admitting two different angular frequencies for a given radius of the circular electrogeodesic. We look both at the weak- and near-field limits of the solution. We use analytic as well as numerical methods in our approach.
\end{abstract}


\vspace{2pc}

\submitto{\CQG}

\section{Introduction}
Study of geodesics has been one of the main tools in the investigation of the physical properties of spacetimes since the very establishment of general relativity---for a general review of both geodesics and electrogeodesics for the Reissner-Nordstr\"{o}m as well as other spacetimes, see, for example, the classical book by Chandrasekhar \cite{chandrasekhar}. Besides revealing the causal structure of the spacetime, geodesics help us build our intuition about the solution and connect it to the classical Newtonian case, if possible. It may also be possible to interpret parameters appearing in the solution and restrict their ranges based on physically reasonable requirements on the behavior of geodesics.

Until this day, new papers are still appearing on geodesics in Kerr \cite{fujita, hod}, which is certainly very important for astrophysical applications, or even in Schwarzschild \cite{kostic}. With the advent of the AdS/CFT conjecture there has been renewed interest in the geodetical structure of spacetimes involving the cosmological constant \cite{cruz, HackmannI, HackmannII}. In these cases, there is usually some test-particle motion which is not allowed in the Schwarzschild spacetime or the interval of admissible radii extends farther, which is also the case here as we can reach the axis. The different allowed ranges could perhaps enable us to distinguish between the various solutions through direct observation. The general method is to find the appropriate effective potential and study its properties and that is also the approach we adopt in the present paper.

The Maxwell field possibly present in the spacetime influences test-particle motion in two ways: firstly by deforming the spacetime, which also has an effect on neutral particles (or even photons \cite{villanueva}), and, secondly, by generating Lorentz force acting upon charged particles. We focus here on the second effect, which was also studied in \cite{bicakI, bicakII} in the case of Kerr-Newmann solution, where there are two different angular frequencies for a circular equatorial orbit of a given radius due to co- and counterrotation of test particles. Papers \cite{olivares} and \cite{grunau} investigate charged electrogeodesics in Reissner-Nordstr\"{o}m with and without the cosmological constant, respectively, but they do not discuss the existence of double frequencies of circular orbits unlike \cite{pugliese} where the two different solutions are discussed for the pure Reissner-Nordstr\"{o}m spherically symmetric spacetime. Circular orbits are important in astrophysics where they determine the appearance of accretion disks around compact objects. It is thus of interest that a spacetime admits orbits of the same radius but different angular velocities. In principle, the double frequencies could show in observations of plasma orbiting charged sources.

In this paper we introduce charged test particles to an axially symmetric static spacetime consisting of two black holes of charges equal to their masses to ensure a static equilibrium between their gravitational attraction and electrostatic repulsion. That is, these are in fact two extremally charged black holes. This is a special case of the Majumdar-Papapetrou class of solutions \cite{majumdar, papapetrou}, which admit point charges with a flat asymptotic region \cite{hartle}. There are two independent scales characterizing the spacetime: it is determined by the masses of the two black holes and their distance but we can always rescale the masses by the distance so there are only two free parameters. We find static positions of the test particles and compare them both to the geodesic case studied previously \cite{wunsch} and the classical non-relativistic solution. We then specifically investigate linear and circular motion of these charged test particles, focussing on analytic solutions while also using numerical approach as required in the study of geodesic deviation.

Although the spacetime is not physically realistic due to its extremal charge, it is an interesting exact solution exhibiting axial symmetry. In fact, with multiple black holes of this kind, one can prescribe any discrete symmetry or even remove it altogether. Perhaps unrealistic but the studied spacetime is still mathematically rewarding since this is an exact solution of Einstein-Maxwell equations and we can use analytic approach throughout most of our paper.

The paper is structured as follows: in Section \ref{Newtonian Case} we review the Newtonian case of two charged point masses in static equilibrium and study motion of charged test particles in their field to be able to check and compare our later results. The background field is the classical analog of the extreme Reissner-Nordstr\"{o}m di-hole metric, which we introduce in Section \ref{The spacetime}. We then investigate static electrogeodesics (Section \ref{Static electrogeodesics}) and test particles oscillating along the $z$-axis (Section \ref{Oscillation along z}). In Section \ref{Circular electrogeodesics} we study general circular electrogeodesics to concentrate on the equatorial case in Section \ref{Circular electrogeodesics within the equatorial plane}. In the final Section \ref{Deviation of geodesics} we briefly look at the stability of geodesics from the point of view of geodesic deviation.

\section{The Newtonian case}\label{Newtonian Case}
Let us start with the Newtonian case of two static massive point charges with their gravitational attraction balanced by their electrostatic repulsion and then let us add a charged massive test particle to study its motion. Suppose that the sources of the field have masses $M_1, M_2$ and charges $Q_1, Q_2$ (of the same sign) chosen in such a way that the particles are in a static equilibrium regardless of their positions. We have the relation (in \texttt{CGS}):
\begin{equation}
|Q_1| = \sqrt{G}M_1, |Q_2| = \sqrt{G}M_2,
\end{equation}
where $G$ is Newton's gravitational constant---in the following we use the geometrized units $G=c=1$.\footnote{For two particles, one can choose any combination of charges and masses satisfying $Q_1 Q_2 = M_1 M_2$. The particular choice $Q_1/M_1 = Q_2/M_2 = 1$ is required for three or more sources of the field and in GR.} Without loss of generality we choose both charges to be positive\footnote{When the charges are negative the electric potential (or the 4-potential in GR) changes sign.} and put the particles to $z=a>0$ ($M_1$) and $z=-a$ ($M_2$), respectively, using the standard cylindrical coordinate system $\rho, \phi, z$\footnote{In fact, we can rescale the problem and measure $\rho, z, t, M_1,$ and $M_2$ in terms of $a$ so that $a=1$.}. The electrostatic and gravitational potentials, $\varphi_e$ and $\varphi_g$, read
\begin{equation}\label{potential}
\varphi_e= \left[\frac{M_1}{\sqrt{\rho ^2+(z-a)^2}}+\frac{M_2}{\sqrt{\rho ^2+(z+a)^2}}\right], \varphi_g = -\varphi_e.
\end{equation}
The Lagrangian $\mathcal{L}$ of the test particle of charge-to-mass ratio $q$ is
\begin{equation}\label{lagrangian}
\mathcal{L} = \frac{1}{2}\left(\dot{\rho}^2+\rho^2 \dot{\phi}^2+\dot{z}^2\right)-\left(\varphi_g+q\varphi_e\right),
\end{equation}
where the dot denotes derivative with respect to Newtonian absolute time while in the relativistic case (from Section \ref{The spacetime} onward) it denotes derivative with respect to proper time. We thus obtain a set of equations
\begin{eqnarray}
\fl 0  &=& \rho  \left\lbrace \dot{\phi }^2+ \left(q- 1\right)\left[\frac{M_1}{\left(\rho ^2+(z-a)^2\right)^{3/2}}+\frac{M_2}{\left(\rho ^2+(z+a)^2\right)^{3/2}}\right]\right\rbrace - \ddot{\rho },\\
\fl 0  &=& - \rho  \left(\rho  \ddot{\phi }+2 \dot{\rho } \dot{\phi }\right),\\
\fl 0  &=& (1-q) \left\lbrace\frac{M_1 (a-z)}{\left[\rho ^2+(z-a)^2\right]^{3/2}}-\frac{M_2 (a+z)}{\left[\rho ^2+(z+a)^2\right]^{3/2}}\right\rbrace - \ddot{z}. \label{z-direction}
\end{eqnarray}
\subsection{Static and circular orbits}
Looking for circular orbits with $\dot{\rho} = 0, \dot{z} = 0$ we have $\ddot{\phi} = 0$ and thus define $\omega \equiv \dot{\phi}$, finding several solutions. Firstly, the static case $\omega = 0, q = 1$ with the test particle located anywhere (we can immediately see from (\ref{lagrangian}) and (\ref{potential}) that this is equivalent to a free particle, generally moving along a straight line at a constant velocity). Additionally, if $\rho=0$, we need to solve (\ref{z-direction}) only to get static points located along the $z$-axis at
\begin{equation}\label{eq:zequklas}
z_{eq}=a \frac{1-\sqrt{\mathcal{M}}}{1+\sqrt{\mathcal{M}}},
\end{equation}
where $\mathcal{M} \equiv M_1/M_2$ and the test particle is allowed to have an arbitrary charge-to-mass ratio $q$.

Finally, we have a solution of the form
\begin{eqnarray}
\mathcal{M} &=& \frac{a+z}{a-z}\left[1-\frac{4 a z}{(a+z)^2+\rho ^2}\right]^{3/2},\\
\omega ^2 &=& \left(1-q\right) \left[\frac{M_1}{\left((z-a)^2+\rho ^2\right)^{3/2}}+\frac{M_2}{\left((a+z)^2+\rho ^2\right)^{3/2}}\right].
\end{eqnarray}
It follows that $|z| < a$ and $1 \geq q$ so the circular orbits may only occur in planes parallel to $z=0$ between the two point masses.
\subsection{Oscillation along the $z$-axis}
It is of interest that there are also periodic solutions lying entirely within the $z$-axis. These necessarily entail the condition $|z|<a$\footnote{In fact, one needs to follow through with (\ref{integrated_classical_z_motion}) outside of the interval $[-a;a]$ to find out there is only a single turning point in $[-\infty;-a]$ and a single turning point in $[a;\infty]$ so that no periodic motion is possible there.}, using which, the equation of motion can be rewritten as
\begin{equation}\label{classical_z_motion}
\frac{1}{1-q}\ddot{z} - \frac{M_1}{\left(a-z\right)^2} + \frac{M_2}{\left(a+z\right)^2} = 0.
\end{equation}
Integrating the last relation, we find
\begin{equation}\label{integrated_classical_z_motion}
    \frac{1}{2} \frac{1}{1-q} \dot{z}^2 - \frac{M_1}{a-z} - \frac{M_2}{a+z} = \mathcal{E} = \mbox{const}.
\end{equation}
We will not give here the full solution as it leads to a complicated expression involving elliptic integrals and, instead, we will simply find the turning points with $\dot{z}=0$:
\begin{equation}\label{classical_turning_points}
    z_{1,2} = \frac{M_1-M_2}{2\mathcal{E}} \left(1 \pm \sqrt{1 + 4a\mathcal{E}\frac{M_1+M_2+a\mathcal{E}}{(M_1-M_2)^2}} \right).
\end{equation}
We conclude that, for $q>1$ there are generally periodic orbits with purely axial motion. It seems that the position of the turning points is independent of the test particle's charge. However, $\mathcal{E} = E/(1-q)$ where $E$ is energy per unit mass of the particle.
\begin{figure}[ht]
\begin{center}
\subfloat[Position.]{\includegraphics[width=0.4\textwidth]{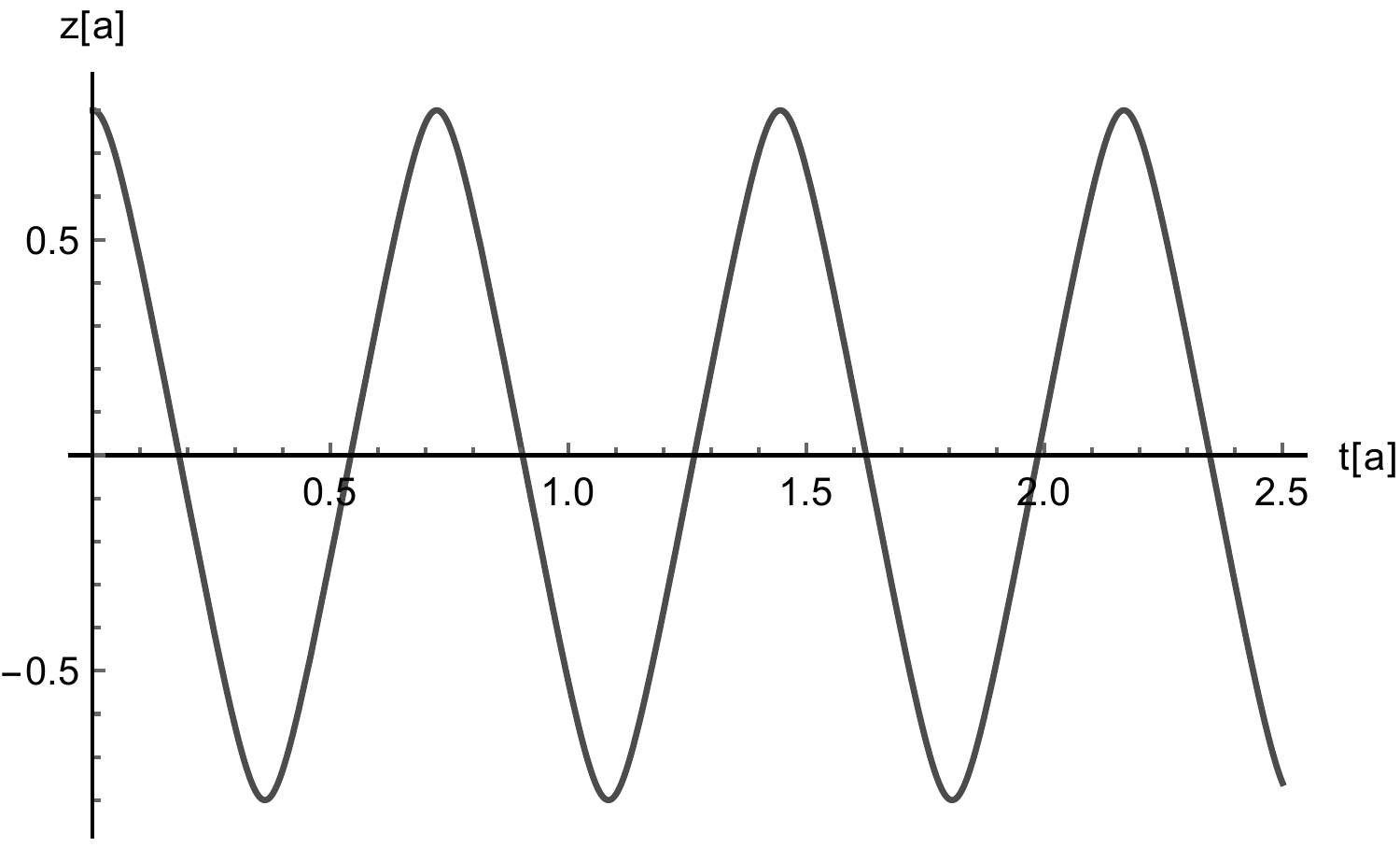}} \hspace{1cm}
\subfloat[Velocity.]{\includegraphics[width=0.4\textwidth]{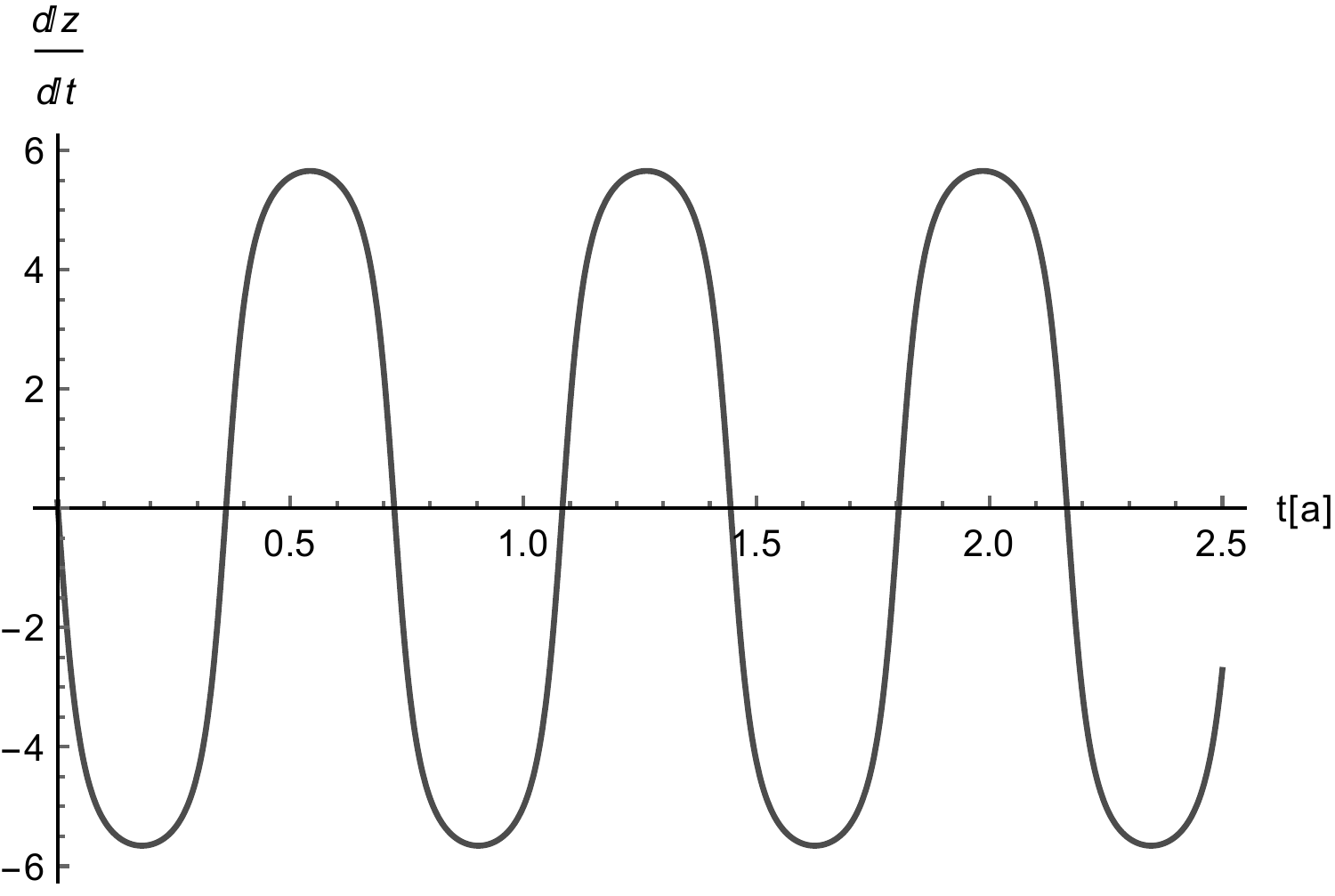}}
\end{center}
\caption{An example of a numerical, periodic Newtonian solution along the $z$-axis with $M_1 = M_2 = a, q=5.5$. The turning points are located at $z_{1,2} \approx \pm 0.8a$. We plot here the position and velocity of a test particle dropped from $z(0)=0.8a$ with an initial velocity $\dot{z}(0)=0.1$.}
\label{tab:oscCLASS}
\end{figure}

We now proceed to discuss the relativistic situation. When discussing the Newtonian limit of the relativistic solutions, we must bear in mind that the above solutions are only relevant for slow motion and weak fields, i.e., at distances from both point sources much greater than their masses.
\section{The extreme Reissner-Nordstr\"{o}m di-hole metric}\label{The spacetime}
The extreme Reissner-Nordstr\"{o}m di-hole metric reads
\begin{equation}\label{eq:MajPapMetrika}
\mathrm{d}s^2 = -U^{-2}\diff{t}^2+U^2\diff{\vec{x}} \cdot \diff{\vec{x}},
\end{equation}
with
\begin{equation}
U(\vec{x}) = 1 + \frac{M_1}{\sqrt{x^2+y^2+\left(z-a\right)^2}} + \frac{M_2}{\sqrt{x^2+y^2+\left(z+a\right)^2}},
\end{equation}
where $\vec{x}$ denotes the spatial position of point sources in Cartesian coordinates. The geometry describes a system consisting of two static black holes with charges equal to their masses and located on the $z$-axis at $z = \pm a$. Their electromagnetic field is described by the 4-potential
\begin{equation}
A = \frac{1}{U} \diff{t}.
\end{equation}
The geometry is axially symmetric so it is useful to replace the Cartesian coordinates with the standard cylindrical ones, $\rho, \phi,z$. The metric, function $U$, and the Maxwell 2-form, $F$, then take the form
\begin{eqnarray}
\mathrm{d}s^2 &=& -U^{-2}\diff{t}^2+U^2(\diff{\rho}^2+\rho^2 \diff{\phi}^2 + \diff{z}^2),\\
U(\rho, z) &=& 1 + \frac{M_1}{\sqrt{\rho ^2 +\left(z-a\right)^2}}+\frac{M_2}{\sqrt{\rho ^2 +\left(z+a\right)^2}},\\
F &=&  \frac{1}{U^2}\left(\pder{U}{\rho}\diff{t} \wedge \diff{\rho}+\pder{U}{z}\diff{t}\wedge\diff{z}\right).
\end{eqnarray}
Consider a test particle with a charge-to-mass ratio $q$. Its Lagrangian reads
\begin{equation}
\mathcal{L} = -\sqrt{-g_{\mu \nu}\dot{x}^{\mu}\dot{x}^{\nu}}+ q\dot{x}^{\sigma}A_{\sigma}=\frac{1}{U}\left[q \dot{t}-\sqrt{\dot{t}^2-U^4 \left(\rho ^2 \dot{\phi }^2+\dot{\rho }^2+\dot{z}^2\right)}\right],
\end{equation}
yielding the following equations of motion
\begin{eqnarray}
\label{eq:elgeodmotion1}
\fl 0  &=& \frac{1}{U}\left[\left(\dot{z}\pdermin{U}{z}+\dot{\rho}\pdermin{U}{\rho}\right)(qU-2\dot{t})+\ddot{t}U\right],\\
\label{eq:elgeodmotion2}
\fl 0  &=& \left(\ddot{\rho}-\rho  \dot{\phi}^2\right)+U^{-5}(qU\dot{t}-\dot{t}^2)\pdermin{U}{\rho}+\frac{1}{U}\left[2\dot{z}\dot{\rho}\pdermin{U}{z}+(\dot{\rho}^2-\rho ^2 \dot{\phi}^2-\dot{z}^2)\pdermin{U}{\rho}\right],\\
\label{eq:elgeodmotion3}
\fl 0  &=& \frac{1}{\rho  U}\left[U \left(2 \dot{\rho} \dot{\phi}+\rho  \ddot{\phi}\right)+2\rho \dot{\phi}\left(\dot{z}\pdermin{U}{z}+\dot{\rho}\pdermin{U}{\rho}\right)\right],\\
\label{eq:elgeodmotion4}
\fl 0  &=& \frac{1}{U^5}\left\{ U^5 \ddot{z}+\left(qU\dot{t}-\dot{t}^2\right)\pdermin{U}{z}+U^4\left[(\dot{z}^2-\rho ^2 \dot{\phi}^2-\dot{\rho}^2)\pdermin{U}{z}+2\dot{z}\dot{\rho}\pdermin{U}{\rho}\right]\right\}.
\end{eqnarray}
The coordinates $t$ and $\phi$ are cyclic so the integrals of motion are
\begin{equation}
E \equiv \pder{\mathcal{L}}{\dot{t}} = \frac{q U-\dot{t}}{U^2}, L_z \equiv \pder{\mathcal{L}}{\dot{\phi}} = \rho ^2 U^2 \dot{\phi }.
\end{equation}
These are the energy of the particle, $E$, and projection of its angular momentum on the $z$-axis, $L_z$. For a vanishing charge of the test particle, these relations are consistent with those for an uncharged test particle \cite{wunsch}.

\section{Static electrogeodesics}\label{Static electrogeodesics}
For simplicity, we first investigate the static solutions $x^{\mu} = \left(t, \rho _0,\phi _0, z _0 \right)$ with spatial coordinates independent of the proper time, $\tau$, to obtain
\begin{eqnarray}
\dot{t}^2=U^2,\ddot{t}&=&0,\\
U^{-5}\left(q U-\dot{t}\right)\dot{t}\pdermin{U}{\rho}&=&0,\\
U^{-5}\left(q U-\dot{t}\right)\dot{t}\pdermin{U}{z}&=&0.
\end{eqnarray}

\begin{figure}[ht]
  \centering
    \includegraphics[width=75mm]{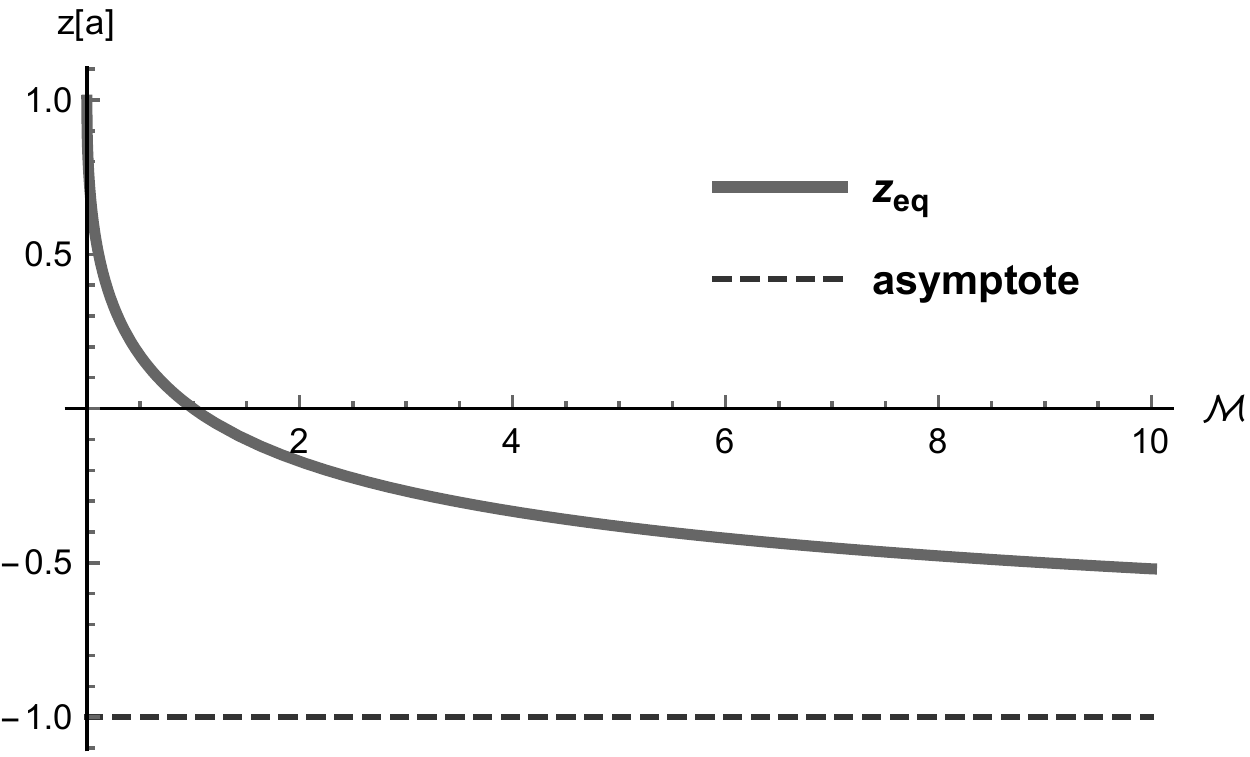}
  \caption{Position of a stationary test particle along the $z$-axis as a function of the black holes' mass ratio $\mathcal{M} = M_1/M_2$.}
  \label{fig:zequ}
\end{figure}

A particularly simple solution with arbitrary $\rho _0, z_0, \phi _0$ reads $x^{\mu} = \left(U \tau, \rho _0,\phi _0,z_0 \right)$ and requires $q=1$. In fact, since the charge-to-mass ratio of the test particle is the same as for the two black holes, this corresponds to an exact solution of the full Einstein-Maxwell equations apart from the field due to this additional point source itself. There are, however, additional static solutions, requiring $\pdermin{U}{\rho}=0$ and $\pdermin{U}{z}=0$ which means there is a single point $z_{eq}$ on the $z$-axis where a test particle can be static. It can have an arbitrary charge since $q$ drops out of the equations. From $ \pdermin{U}{z} = 0$ we infer
\begin{equation}
\mathcal{M} \equiv \frac{M_1}{M_2}=\left(\frac{a-z}{a+z}\right)^2,
\end{equation}
with the solution
\begin{equation}\label{eq:zequrov}
z_{eq}=a \frac{1-\sqrt{\mathcal{M}}}{1+\sqrt{\mathcal{M}}}.
\end{equation}
This relation, plotted in Fig. \ref{fig:zequ}, is identical to the Newtonian expression (\ref{eq:zequklas}). For equal-mass black holes, the stationary point is located at the center while for $\mathcal{M} \neq 1$ it lies closer to the less massive black hole and it is identical to that of an uncharged particle as calculated in \cite{wunsch}.

\section{Oscillation along the $z$-axis}\label{Oscillation along z}
\begin{figure}[ht]
\begin{center}
\subfloat[Position.]{\includegraphics[width=0.4\textwidth]{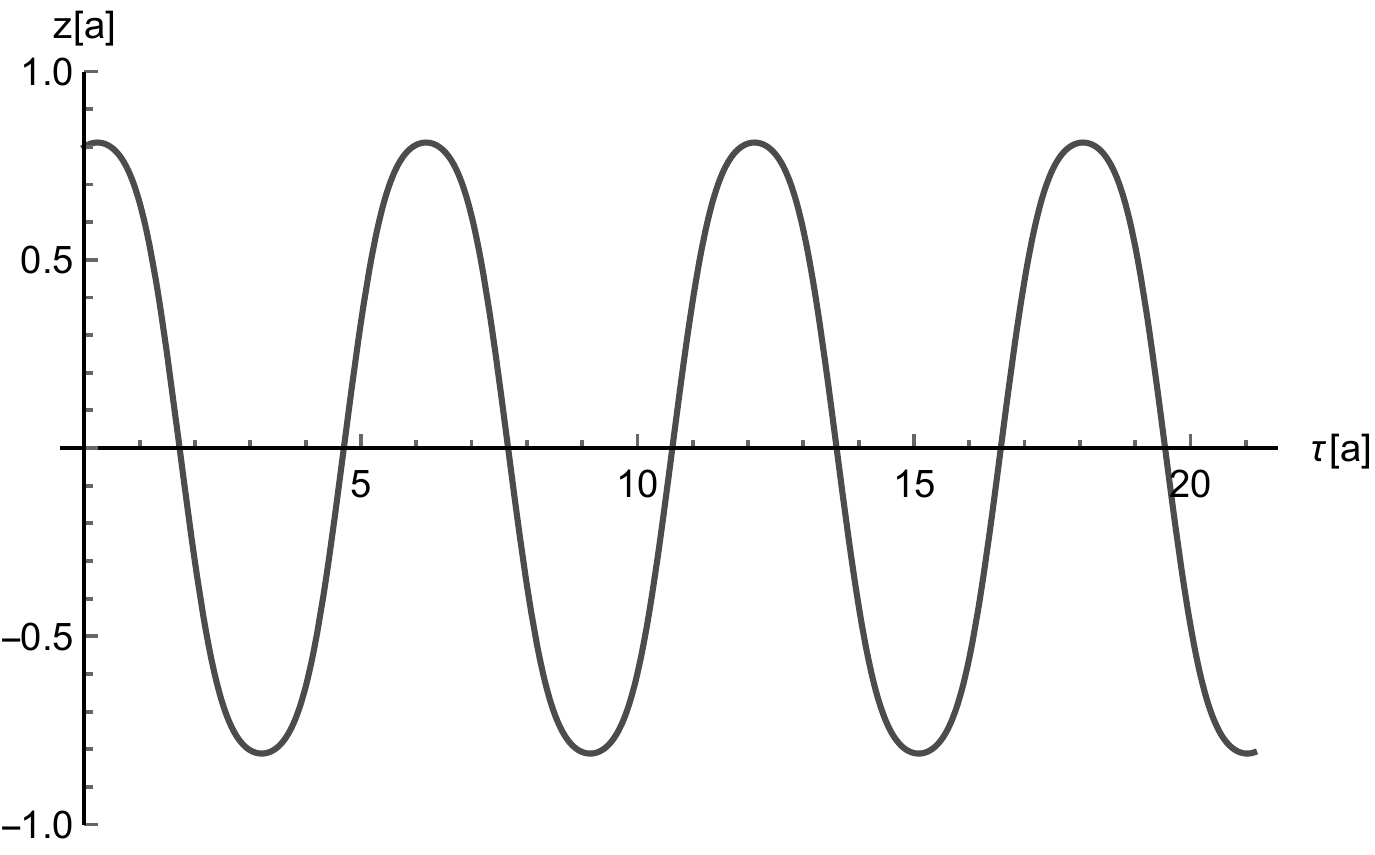}} \hspace{1cm}
\subfloat[Velocity.]{\includegraphics[width=0.4\textwidth]{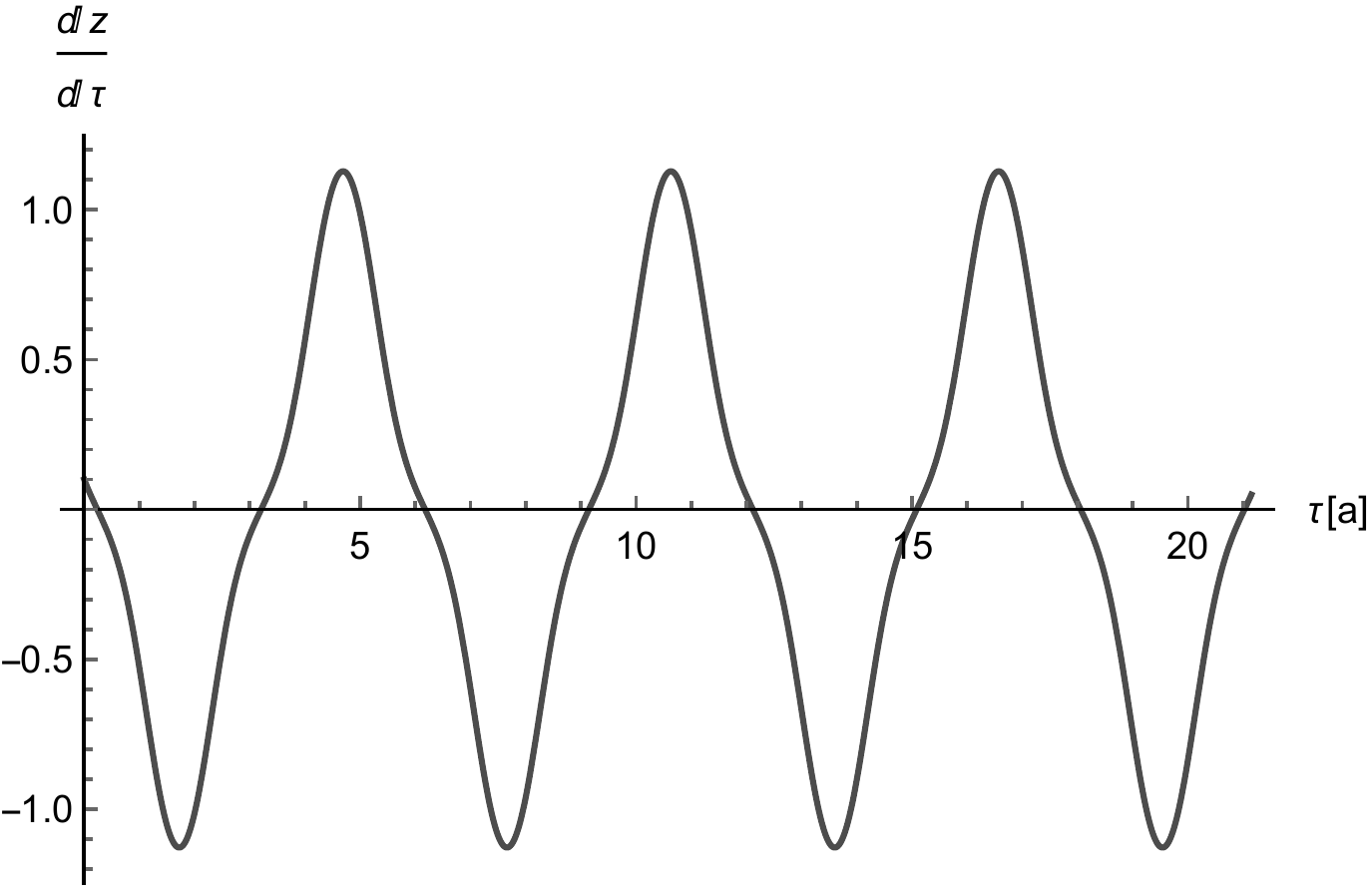}}
\end{center}
\caption{An example of a periodic numerical solution along the z-axis with $M_1 = M_2 = a, q=5.5,E \approx 0.66, z(0)=0.8a,\dot{z}(0) > 0$. The turning points are located at $z_{1,2} \approx \pm 0.81a$. We plot here the position and velocity of the test particle as functions of its proper time. We notice the frequency is about ten times lower than in Figure~\ref{tab:oscCLASS}.}
\label{tab:oscGR}
\end{figure}

Similarly to the classical case, there are periodic solutions limited to the $z$-axis when the equations of motion reduce to the conservation of energy and normalization of the 4-velocity as follows
\begin{eqnarray}
\dot{t} - U\left(q-U E\right) &=& 0,\\
-\left(q - E U\right)^2+U^2 \dot{z}^2 +1 & = & 0.
\end{eqnarray}
Unlike the classical case, the test particle can reach the non-singular points $z=\pm a$ now and cross the horizon. This, however, does not produce a periodic solution since the particle can never come back. As in the classical case, we need to discuss the ranges $z \in (-\infty;-a]$, $z \in [-a;a]$, and $z \in [a;-\infty)$ separately. We define
\begin{equation}\label{definition of f below}
    f(z) \equiv U^2\dot{z}^2 = -1 + \left(q - E U\right)^2 \geq 0
\end{equation}
and search for the turning points with $f=0$. We always have $\lim_{z \rightarrow \pm a}f(z) = +\infty$. We are thus looking for neighboring turning points separated by an interval with $f>0$ which means we are looking for local maxima. We now rescale the two masses and $z$ by $a$ so that $\tilde{M}_1 \equiv M_1/a$, $\tilde{M}_2 \equiv M_2/a$, and $\tilde{z} \equiv z/a$ and drop the tildes. Let us begin with $z \in (-\infty;-1)$ where we can write
\begin{equation}\label{below rescaled}
    f = -1 + \left[q - E \left(1+\frac{M_1}{1-z} - \frac{M_2}{1+z}\right) \right]^2.
\end{equation}
We find
\begin{equation}\label{derivative below}
    \frac{\mathrm{d}f}{\mathrm{d}z} = -2 E \left(\frac{M_1}{(1-z)^2} + \frac{M_2}{(1+z)^2}\right) \left[q - E \left(1+\frac{M_1}{1-z} - \frac{M_2}{1+z}\right) \right].
\end{equation}
The first bracket never vanishes while the second has no extremum as its derivative yields again the first bracket. Therefore, the second bracket has at most one root within $(-\infty;-1]$ and $f$ has a single extremum here, which is then necessarily a minimum due to the fact it diverges to $+\infty$ at $z=-1$. We thus cannot have periodic motion of test particles here. The same line of reasoning also applies to the case $z \in [1;-\infty)$ so that we now proceed directly to the case $z \in [-1;1]$. We have
\begin{equation}\label{between rescaled}
    f = -1 + \left[q - E \left(1+\frac{M_1}{1-z} + \frac{M_2}{1+z}\right) \right]^2
\end{equation}
and
\begin{equation}\label{derivative between}
    \frac{\mathrm{d}f}{\mathrm{d}z} = -2 E \left(\frac{M_1}{(1-z)^2} - \frac{M_2}{(1+z)^2}\right) \left[q - E \left(1+\frac{M_1}{1-z} + \frac{M_2}{1+z}\right) \right].
\end{equation}
The first bracket has two roots now but only one of them lies within $[-1;1]$. Similarly, the second bracket only has a single local extremum here and thus at most two roots. We conclude that $f$ has at most three local extrema yielding a single possible interval for periodic motion around the middle value. To summarize, oscillatory $z$-motion of test particles is only allowed between the hypersurfaces $z= \pm a$. In \Fref{tab:oscGR} we give an example of such motion.

Investigating now (\ref{between rescaled}), we can express the turning points as follows
\begin{eqnarray}\label{relativistic_turning_points}
z_{k} &=& \frac{E\Delta+ (-1)^{k+1} \sqrt{E^2\Delta^2 + 4 a q_k\left[E\Sigma +a q_k\right]}}{2q_k}, k = 1 \ldots 4,\\
\Delta & \equiv & M_1- M_2 , \Sigma  \equiv  M_1+ M_2, q_{1,2}=E-q+1,q_{3,4}=E-q-1.
\end{eqnarray}
As $\lim_{z \rightarrow \pm 1} f = \infty$, the condition for oscillation is that all the turning points must lie between the black-hole horizons $z_i^2 < 1$ and $f(z_{eq}) > 0$, which is not always the case. This is generally a complicated system so we restrict the situation to black holes of equal masses. We find that $z_3<z_1, z_2<z_4, z_2=-z_1,z_4=-z_3$. The oscillation takes place between the two turning points nearest to the origin. \Tref{pokus} sums up the two possible cases.
\begin{table}[h]
\begin{tabular}{|l|l|l|p{3cm}|}
\hline
Range of $q$ & Range of $E$ & Range of $M$ & Turning points \\ \hline
$q > 1$ & $0 < E < q-1$ & $ 2EM < q-1-E$ & $z_1<z_2$\\ \hline
$q < -1$ & $1+q < E < 0$ & $ 2EM >1-q+E$ & $z_4<z_3$\\ \hline
\end{tabular}
\caption{Regions of parameters where we can have $z$-axis oscillations for black holes of equal masses, $M_1 = M_2 \equiv M$.}
\label{pokus}
\end{table}

In the limit $M/a \rightarrow 0$ all the turning points approach $z= \pm 1$ where their distance from the singularity is of the same order as the mass of the black holes and, therefore, there is no Newtonian limit and we cannot compare (\ref{relativistic_turning_points}) to (\ref{classical_turning_points}).
\section{Circular electrogeodesics}\label{Circular electrogeodesics}
We now assume $\rho, z$ constant so that the normalization and electrogeodesic equations read
\begin{eqnarray}
&&\rho ^2 U^4 \dot{\phi}^2 -\dot{t}^2=-U^2,\\
&&\label{eq:kruhcas} \ddot{t} = 0,\\
&&\label{eq:rhocas}\left(qU-\dot{t}\right)\dot{t}\pdermin{U}{\rho}-\rho\dot{\phi}^2 U^4 \left(U+\rho\pdermin{U}{\rho}\right)=0,\\
&&\label{eq:uhelcas}\ddot{\phi} = 0,\\
&&\label{eq:deUdez}\left(-qU\dot{t}+\dot{t}^2+\rho ^2 U^4 \dot{\phi}^2\right)\pdermin{U}{z}=0.\label{}
\end{eqnarray}
Equations (\ref{eq:kruhcas}, \ref{eq:uhelcas}) can be easily integrated to yield $\phi = \omega \tau$ and $t = \gamma \tau$. If we require the bracket in (\ref{eq:deUdez}) to vanish and combine this with (\ref{eq:rhocas}), we only get the previous static solution with $q=1$. Therefore, we must have $\pdermin{U}{z} = 0$, which results in the following relation
\begin{equation}\label{eq:kruhprvni}
\mathcal{M} \equiv \frac{M_1}{M_2} =\frac{a+z}{a-z} \left[1-\frac{4az}{\rho ^2+(z+a)^2}\right]^{3/2},
\end{equation}
which can be inverted (for $z \not = 0$, see the next section) as
\begin{equation}
\rho \left(z\right)^2 = \frac{4 a z}{1-\mathcal{M}^{2/3} \left(\frac{a-z}{a+z}\right)^{2/3}}-(a+z)^2.\label{orbital radius}
\end{equation}
As in the Newtonian case, these orbits only exist for ${|z|<a}$. Moreover, they only admit ${z \geq \frac{\mathcal{M}-1}{\mathcal{M}+1}} \geq 0$ (${\rho \rightarrow \infty}$ for equality) and ${z \leq z_{eq} = \frac{1-\sqrt{\mathcal{M}}}{1+\sqrt{\mathcal{M}}}} \leq 0$ (${\rho = 0}$ for equality). Finally, we get the following formulae for $\omega$ and $\gamma$
\begin{eqnarray}
\label{eq:kvadr1} \gamma^2 &=& \rho ^2 U^4 \omega ^2+U^2,\\
\label{eq:kvadr2} \omega ^2\rho U^4\left(U+\rho \pdermin{U}{\rho}\right) &=& \gamma\pdermin{U}{\rho}\left(qU-\gamma\right).
\end{eqnarray}
Since these equations are generally quadratic, we expect up to two solutions for $\omega$ and $\gamma$ (the sign of $\omega$ only describes clockwise or counterclockwise motion). However, (\ref{eq:kvadr1}) and (\ref{eq:kvadr2}) become linear for the special case of $2 \rho \pdermin{U}{\rho} + U=0$, yielding
\begin{eqnarray}
\label{eq:special_case_gamma} \gamma&=&\frac{U}{q},\\
\label{eq:special_case_omega}\omega &=&\frac{\sqrt{1-q^2}}{q\rho U},
\end{eqnarray}
with $0 < q \leq 1$. This gives a particular set of radii for possible orbits which, interestingly, coincide with the positions of null circular geodesics as discussed in \cite{wunsch} so we have the same radius with a photon or a charged massive particle (at different velocities, of course).

On the other hand, the general solution of (\ref{eq:kvadr1}) and (\ref{eq:kvadr2}) reads
\begin{eqnarray}
\label{eq:kruhB} \gamma_{\pm}&=&U\frac{q \rho \pdermin{U}{\rho} \pm  \sqrt{\left(q^2+8\right)\left(\rho\pdermin{U}{\rho}\right)^2+12 U\rho\pdermin{U}{\rho}+4 U^2}}{2 (2 \rho \pdermin{U}{\rho} + U)},\\
\label{eq:kruhomegapm}\omega_{\pm} ^2 &=&\frac{\gamma_{\pm}^2-U^2}{\rho ^2 U^4}.
\end{eqnarray}
We conclude that there can be two different values of angular velocity for a given orbital radius since the electrogeodesic equation contains both linear and quadratic terms in~$\gamma$ with the linear one stemming from the Lorentz force. This is very different from the case of a neutral test particle and also from the classical Newtonian case with a charged particle where in both cases the orbital radius determines a single angular velocity.

In the following discussion we specialize to trajectories within the equatorial plane. This fixes $z=0$ but, on the other hand, $\rho$ can be arbitrary and we ultimately have 3 independent parameters. If, instead, we left the equatorial plane then $\rho$ would be fixed by (\ref{orbital radius}) but the masses of the two holes would be independent, resulting in 4 parameters which would render the discussion even more complicated without bringing any new type of solutions into the play and we thus only investigate the simpler case of equatorial orbits.

\section{Circular electrogeodesics within the equatorial plane}\label{Circular electrogeodesics within the equatorial plane}
We now investigate the special case of $z = 0$. It follows from (\ref{eq:kruhprvni}) that this is only possible if the two black holes have equal masses, $M_1 = M_2 \equiv M$. The function $U$ then simplifies to
\begin{equation}
U(\rho) = 1 + \frac{2 M}{\sqrt{\rho ^2+a^2}}.
\end{equation}
In this case, the orbital radius, $\rho$, can be arbitrary as (\ref{eq:kruhprvni}) is an identity and (\ref{orbital radius}) does not apply.

\subsection{Weak- and near-field limits}
In the asymptotically flat region $M/\rho \rightarrow 0$ (the Newtonian limit) we find
\begin{eqnarray}
\gamma_{\pm} \approx \pm 1 + O\left(\frac{M}{\rho}\right),\\
\rho^2 \omega_{\pm}^2 \approx 2M\frac{1\mp q}{\rho} + O\left(\left[\frac{M}{\rho}\right]^2\right).
\end{eqnarray}
and near the axis with $\rho/a \rightarrow 0$ (strong relativistic effects if $M \not \ll a$) we have
\begin{eqnarray}
\gamma_{\pm}&\approx & \pm \left(1+\frac{2 M}{a}\right) + O\left(\left[\frac{\rho}{a}\right]^2\right),\\
a^2 \omega_{\pm}^2 &\approx & \frac{2 \frac{M}{a} (1\mp q)}{(1+2 \frac{M}{a})^3} + O\left(\left[\frac{\rho}{a}\right]^2\right).
\end{eqnarray}
As we require ${\gamma>0}$, we only have a single solution with the upper sign in both asymptotic regions. The leading order of the angular velocity requires $q<1$ and it is consistent with the classical solution (near the axis we need to assume a weak field with $M\ll a$). Therefore, asymptotically, a given radius of orbit only corresponds to a single orbital frequency. This, however, is not the case generally as we will discuss in the following section. This might have an observable effect for particles orbiting compact charged objects. It is of interest that this does not occur in the strongest field along the axis but farther outside.
\subsection{Existence of solutions}
We proceed by discussing the regions where both solutions (\ref{eq:kruhB}), (\ref{eq:kruhomegapm}) exist. This also occurs in Reissner-Nordstr\"{o}m geometry as discussed in \cite{pugliese} and mentioned in \cite{grunau}. Our requirements are that $\gamma_{\pm}$ be positive and $\omega_{\pm}^2$ non-negative. \Eref{eq:kruhomegapm} can be written as
\begin{equation}
\omega_{\pm} ^2 =\frac{\left(\gamma_{\pm}-U\right)\left(\gamma_{\pm}+U\right)}{\rho ^2 U^4}
\end{equation}
and we thus have a stronger condition $\gamma_{\pm} \geq U$ since $U$ and $\gamma_{\pm}$ are positive. The discussion is very complicated since we have 3 independent parameters appearing in our expressions (after rescaling everything by $a$). For this reason, we just summarize the results in three tables below.
\begin{table}[ht]
\centering
$
\begin{array}{|l|l|l|}
\hline
\mathrm{Range\ of\ } \rho & \mathrm{Range\ of\ } q & \mathrm{Range\ of\ } M \\ \hline
0<\rho\leq a & q<1 & \mbox{any }M \\ \hline
\rho >a & q<1 & M<M_{lim} \\ \hline
a<\rho <\sqrt{2} a & 0< q<q_{lim} & M_{lim}<M\leq M_{+} \\ \hline
a<\rho <\sqrt{2} a & q_{lim} \leq q<1 & M_{lim}<M \\ \hline
\rho \geq\sqrt{2} a & 0<q<1 & M_{lim}<M\leq M_{-} \\ \hline
\end{array}
$
\caption{Summary of conditions for the `$+$' solution (\ref{eq:kruhB}), (\ref{eq:kruhomegapm}) to exist. }
\label{table:variantaplus}
\end{table}
\begin{table}[ht]
\centering
$
\begin{array}{|l|l|l|}
\hline
\mathrm{Range\ of\ } \rho & \mathrm{Range\ of\ } q & \mathrm{Range\ of\ } M \\
\hline
a<\rho \leq \sqrt{2} a & q_{lim} \leq q & M_{lim}<M \\ \hline
a<\rho \leq \sqrt{2} a & 0<q<q_{lim} &  M_{lim}<M\leq M_{+} \\ \hline
\rho >\sqrt{2} a & q=q_{lim} &  M_{lim}<M\leq \frac{1}{2}M_{lim,2} \\ \hline
\rho >\sqrt{2} a & q<1 \land q \neq q_{lim} &  M_{lim}<M\leq M_{-} \\ \hline
\rho >\sqrt{2} a & q \geq 1 &  M_{lim}<M< M_{lim,2} \\ \hline
\end{array}
$
\caption{Summary of conditions for the `$-$' solution (\ref{eq:kruhB}), (\ref{eq:kruhomegapm}) to exist. }
\label{table:variantaminus}
\end{table}
\begin{table}[ht]
\centering
$
\begin{array}{|l|l|l|}
\hline
\mathrm{Range\ of\ } \rho & \mathrm{Range\ of\ } q & \mathrm{Range\ of\ } M \\
\hline
a<\rho <\sqrt{2} a  & q_{lim}\leq q<1 & M_{lim}<M \\ \hline
a<\rho <\sqrt{2} a  & 0<q<q_{lim} & M_{lim}<M\leq M_{+} \\ \hline
\rho \geq \sqrt{2}a & 0<q<1 & M_{lim}<M\leq M_{-} \\ \hline
\end{array}
$
\caption{Both solutions exist.}
\label{table:varaiantaobe}
\end{table}

The values used in the tables are defined as follows
\begin{equation}
M_{lim} = \frac{1}{2} \sqrt{\frac{\left(\rho ^2+a^2\right)^3}{\left(\rho ^2-a^2\right)^2}}, M_{lim,2} =  \sqrt{\frac{\left(\rho ^2+a^2\right)^3}{\left(\rho ^2-2a^2\right)^2}}, q_{lim} = \frac{ 2 a \sqrt{\rho ^2-a^2}}{\rho ^2},
\end{equation}
\begin{equation}
\fl M_{\pm} = \sqrt{\frac{\left(a^2+\rho ^2\right)^3 \left[\left(2-q^2\right) \rho ^4\pm 2 \rho ^2 \sqrt{\left(1-q^2\right) \left(2 a^2-\rho ^2\right)^2}+4 a^2 \left(a^2-\rho ^2\right)\right]}{\left(4 a^4-4 a^2 \rho ^2+q^2 \rho ^4\right)^2}}.
\end{equation}
These are not constants---as opposed to $M$---but rather functions of $\rho$, $a$, and possibly $q$ (in fact, we can again rescale everything using $a$ as our basic unit). We give below plots of these functions to understand the conditions appearing in the tables. As all the $M$'s diverge for $\rho \rightarrow \infty$ we only get a limited range of admissible radii apart from the second row in \Tref{table:variantaplus}. If $M=M_{lim}$ we get the special case discussed in (\ref{eq:special_case_gamma}, \ref{eq:special_case_omega}).
\begin{figure}[ht]
\begin{center}
\subfloat[$M_{lim}$ and $M_{lim,2}$.]{\includegraphics[width=0.3\textwidth]{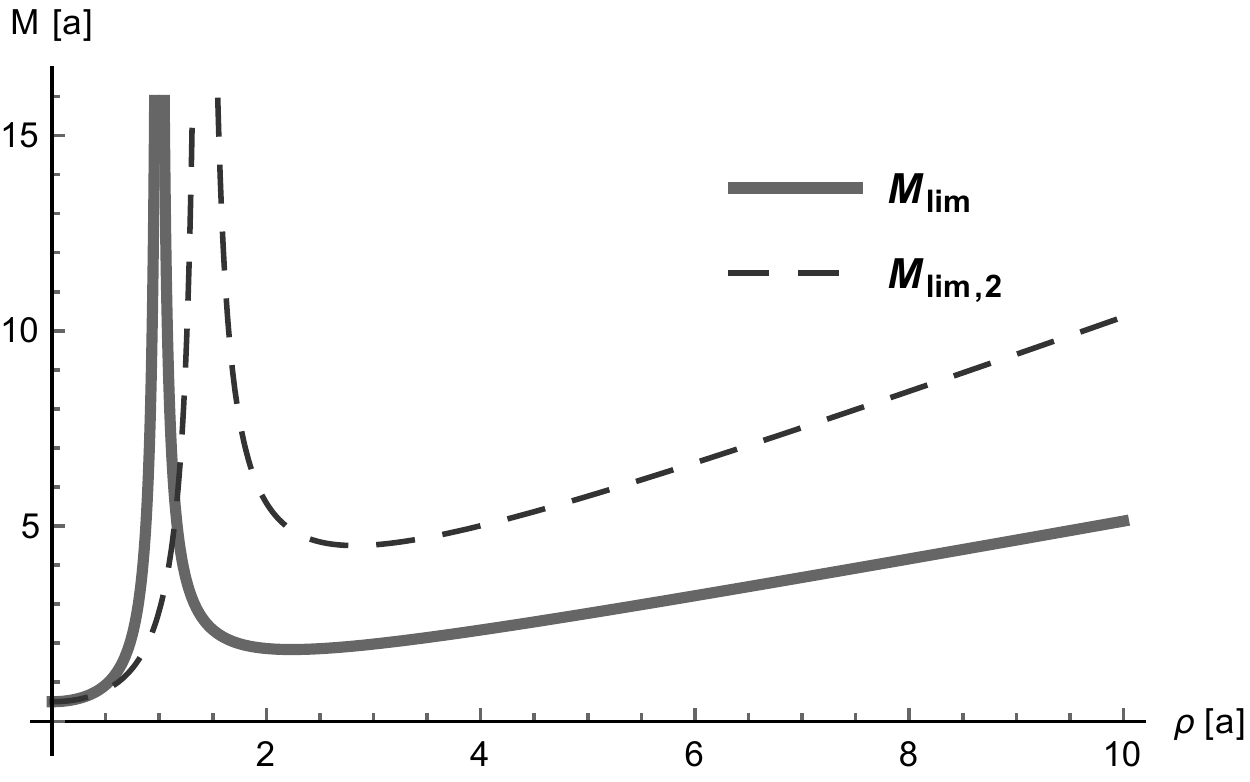}}  \hspace{0.5cm}
\subfloat[$M_{+}$.]{\includegraphics[width=0.3\textwidth]{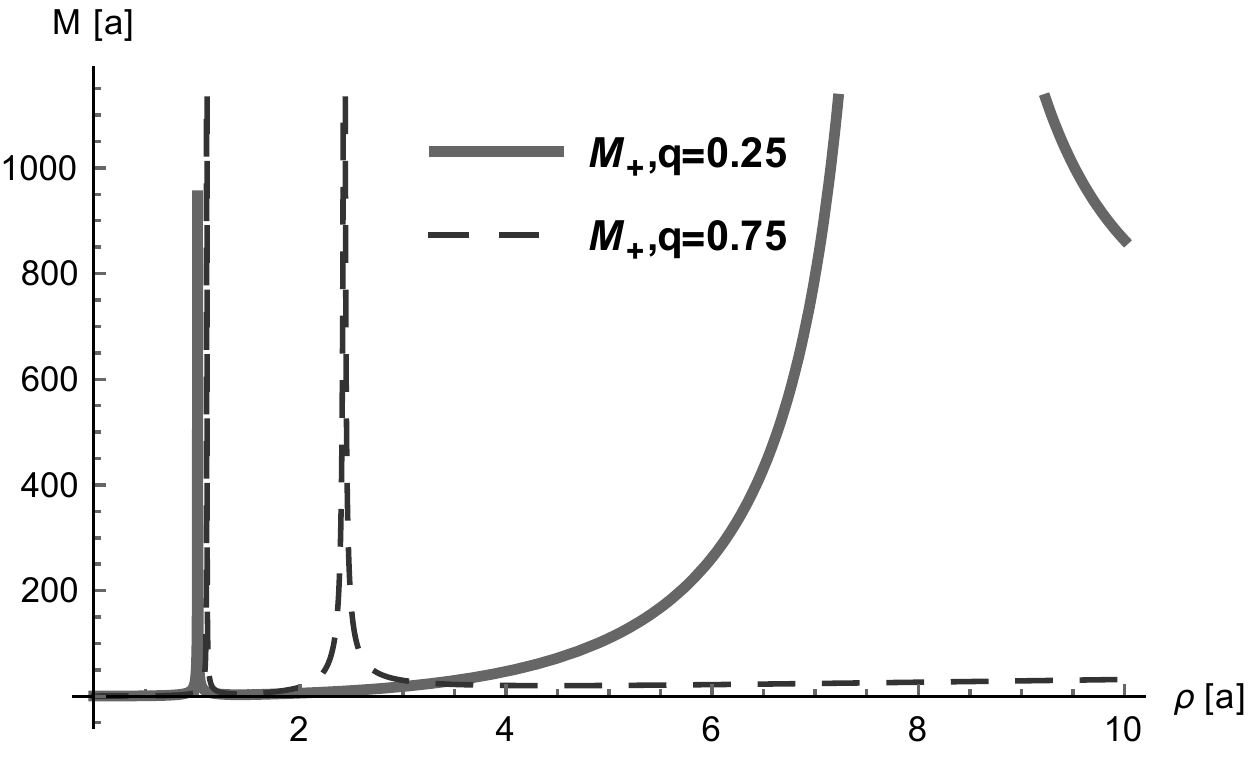}} \hspace{0.5cm}
\subfloat[$M_{-}$.]{\includegraphics[width=0.3\textwidth]{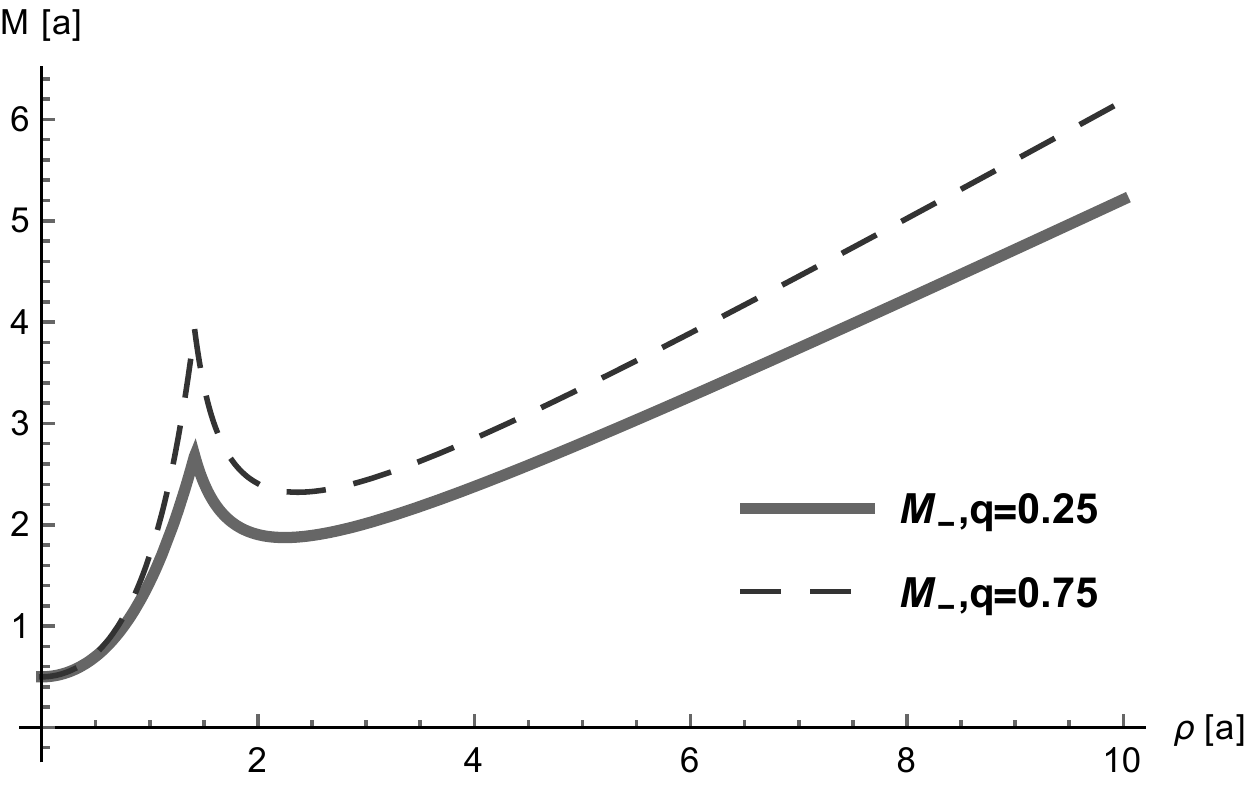}}
\end{center}
\caption{Plots of the functions determining the existence of the two kinds of electrogeodesics. $M_{lim}$ and $M_{lim,2}$ are only functions of $\rho$ while $M_{\pm}$ also include the charge of the test particle.}
\label{tab:XYZ}
\end{figure}

The most interesting option is summarized in \Tref{table:varaiantaobe}. These are the regions in the parameter space where we get both solutions and, therefore, two different frequencies for the same radius of a circular orbit. This, however, can only occur between $\rho=a$ and a finite radius given by the last column in \Tref{table:varaiantaobe}. We thus conclude that we can only get both solutions simultaneously in the vicinity of the axis but not directly upon it.
\section{Deviation of geodesics}\label{Deviation of geodesics}
To examine the stability of circular orbits of neutral test particles and thus, e.g., the possible existence of accretion disks in this system, we now investigate the geodesic deviation. We assume equatorial trajectories and study numerically the evolution of a ring of particles centered on the exact circular geodesic. The general equation of geodesic deviation is
\begin{equation}
\de{\ddot{x}^{\mu}} + \Gamma^{\mu}_{~\alpha \beta}\de{\dot{x}}^{\alpha}u^{\beta} + R^{\mu }_{~\sigma \alpha \beta } u^{\sigma } \de{x} ^{\alpha } u^{\beta } =0,
\end{equation}
with $\de{x}^{\alpha}$ the deviation from the central geodesic. In our case this now yields a set of four equations as follows
\begin{eqnarray}
\fl U^2 \de{\ddot{t}}- \gamma \omega \rho U_{,\rho} \left(  \rho   U_{,\rho}  + U  \right) \de{\phi} + \rho \omega ^2 U_{,\rho} \left( \rho  U_{,\rho}+  U  \right) \de{t} -\gamma  U U_{,\rho} \de{\dot{\rho}} =0, \label{geodevT}\\
\nonumber
\fl \left[U_{,\rho}^2 \left(3 \gamma ^2 +\rho ^2 U^4 \omega ^2 \right) -\rho  U^5 \omega ^2 \left(U_{,\rho}+\rho U_{,\rho \rho}\right)-\gamma ^2 U U_{,\rho \rho} \right]\de{\rho} -\rho \omega U^5 \left(\rho U_{,\rho} +  U \right)\de{\dot{\phi}}+\\
\fl +U \left( U^5\de{\ddot{\rho}} -\gamma  U_{,\rho} \de{\dot{t}} \right) =0, \label{geodevRho}\\
\fl \gamma ~\de{\ddot{t}}-\rho ^2 \omega U^4 \de{\ddot{\phi}} = 0, \label{geodevPhi}\\
\fl \de{\ddot{z}}=0. \label{geodevZ}
\end{eqnarray}
Here, we already used the equation of a circular geodesic and combined equations for $\de{\phi}$ and $\de{t}$. Let us further assume $\de{z} \equiv 0$, which means the ring stays within the equatorial plane $z=0$. We investigate numerically the deformation of a ring with zero initial velocities in both $\phi$ and $\rho$ directions. We thus parametrize the initial conditions as
\begin{equation}
\fl {\de{\rho} (0)\!=\!\sqrt{\rho ^2+\mathcal{R}^2+2 \rho  \mathcal{R} \cos \alpha }-\rho,} \: {\de{\phi} (0)\!=\!\arctan\left(\rho + \mathcal{R} \cos \alpha, \mathcal{R} \sin \alpha \right)\!,} \: {\alpha \in \left[0,2\pi\right)\!,}
\end{equation}
which defines a circle in cylindrical coordinates with radius $\mathcal{R}$, centered at a distance $\rho$ from the origin along the $\rho$ axis. The remaining initial deviation values are set to zero. To depict deformations of the ring, we first transform the resulting deviations to a non-rotating frame ${\de{x},\de{y}}$---with the $\de{x}$ axis along the $\rho$ axis at zero proper time---and then to a co-rotating Cartesian frame ${\de{X},\de{Y}}$. The new deviations are defined as
\begin{eqnarray}
\fl \de{x} \equiv \left( \rho + \de{\rho} \right) \cos \left( \omega \tau + \de{\phi} \right) - \rho \cos \left( \omega \tau \right)\!, \: \de{y} \equiv \left( \rho + \de{\rho} \right) \sin \left( \omega \tau + \de{\phi} \right) - \rho \sin \left( \omega \tau \right)\!,\\
\fl \de{X} \equiv \de{x} \cos \left( \omega \tau \right) + \de{y} \sin \left( \omega \tau \right)\!, \: \de{Y} \equiv \de{y} \cos \left( \omega \tau \right) - \de{x} \sin \left( \omega \tau \right)\!.
\end{eqnarray}
Thus, $\de{X}$ represents locally the $\rho$ direction and $\de{Y}$ the $\phi$ direction at each point along the central geodesic.
\begin{figure}[ht]
\begin{center}
\subfloat[$M=1a, \rho = 0, \mathcal{R}=0.08a$]{\includegraphics[width=0.4\textwidth]{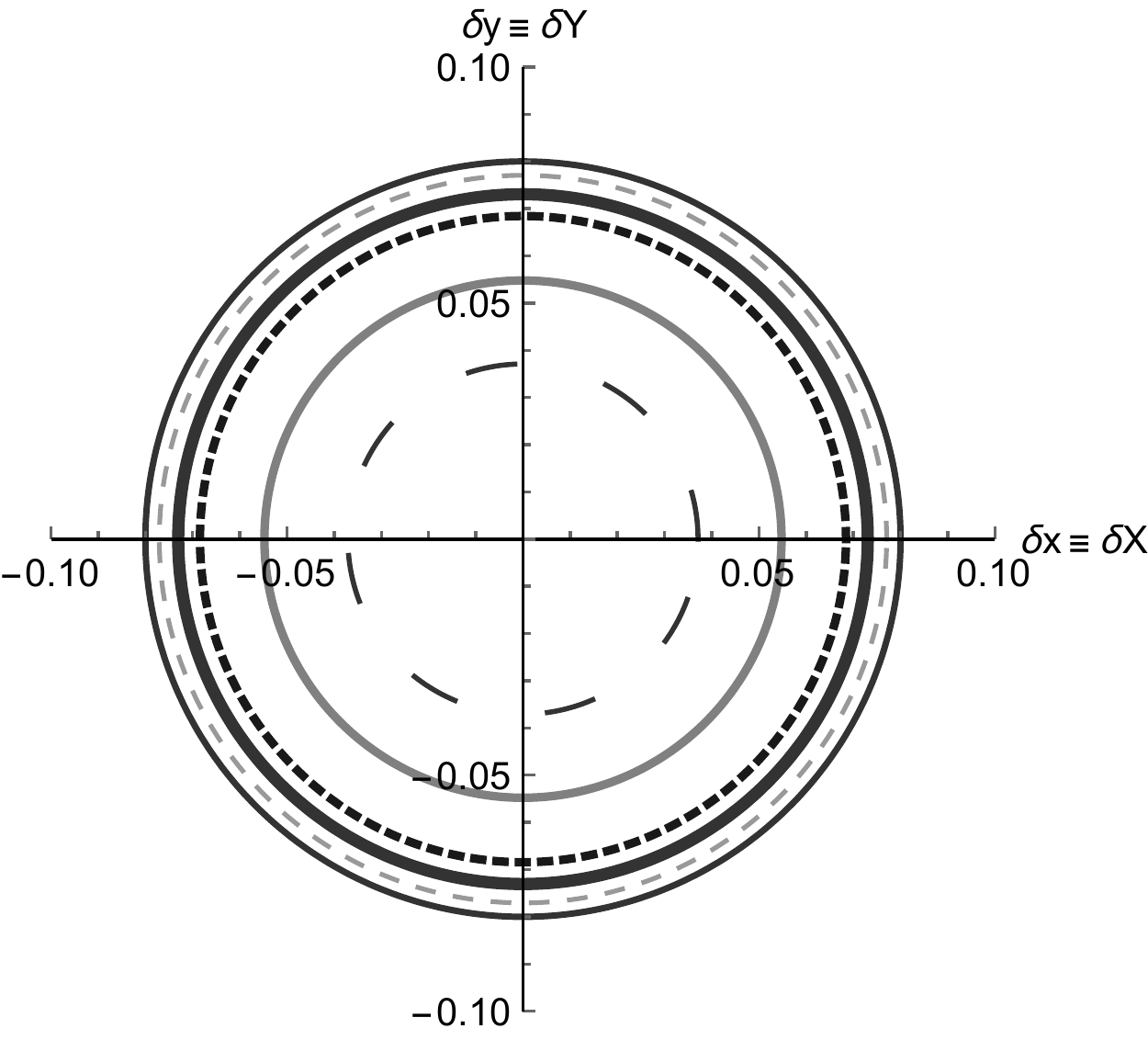}} \hspace{1cm}
\subfloat[$M= 4 a, \rho = 0.3 a, \mathcal{R}=0.008 a$]{\includegraphics[width=0.4\textwidth]{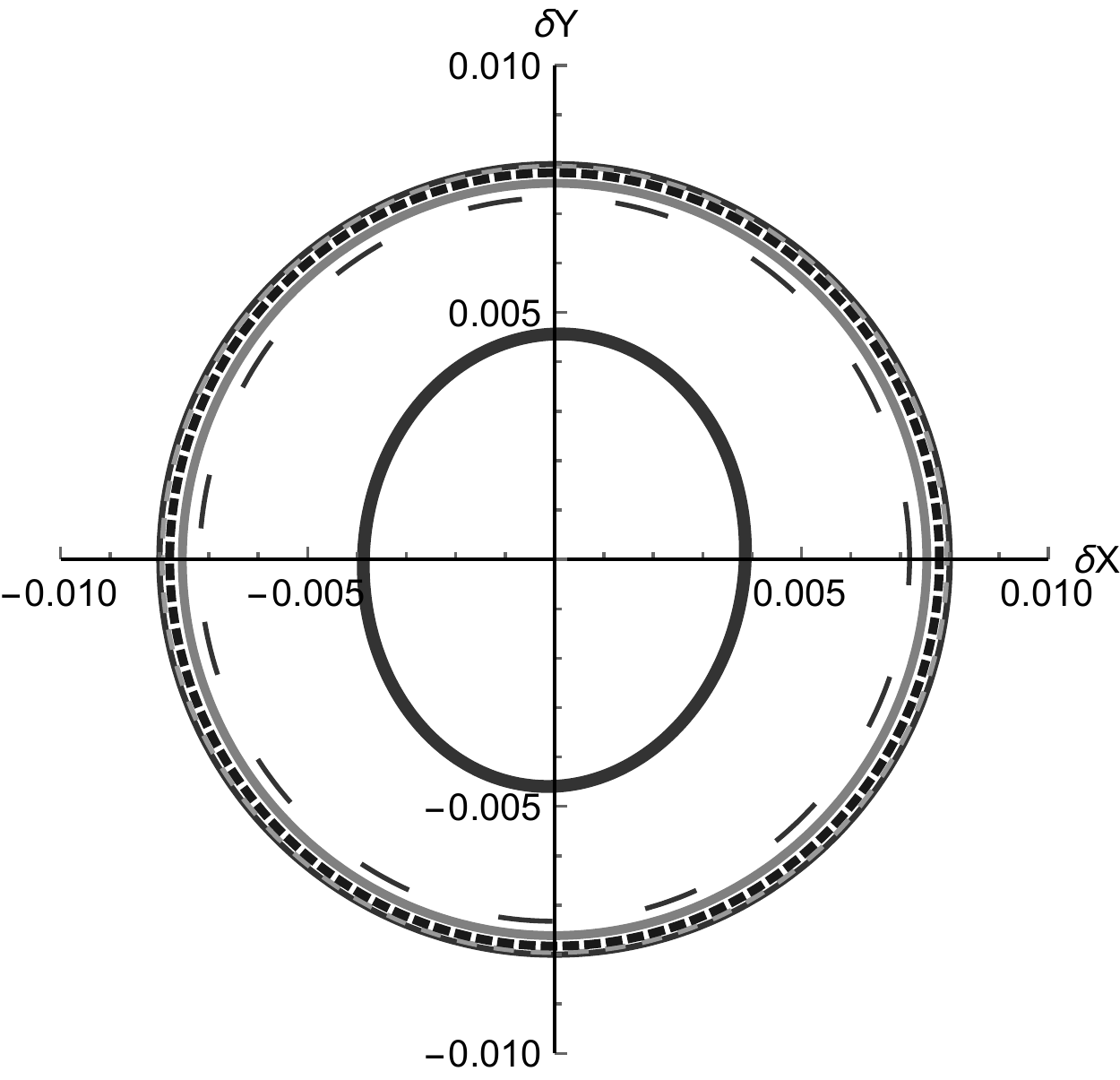}}\\
\subfloat[$M= 4 a, \rho = 1.2 a, \mathcal{R}=0.01 a$]{\includegraphics[width=0.4\textwidth]{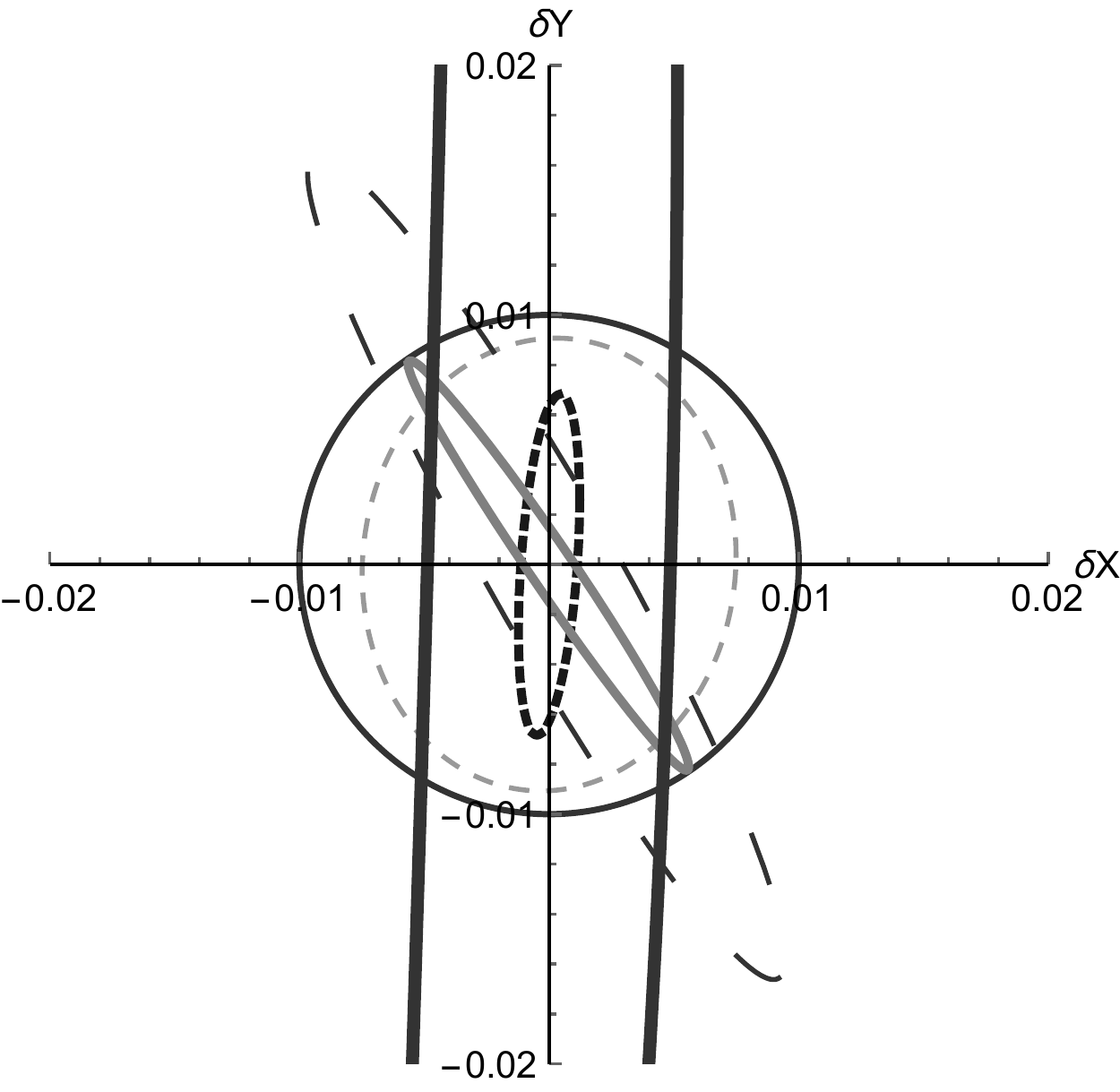}} \hspace{1cm}
\subfloat[$M= 3 a, \rho = 5.81 a, \mathcal{R}=0.01 a$]{\includegraphics[width=0.4\textwidth]{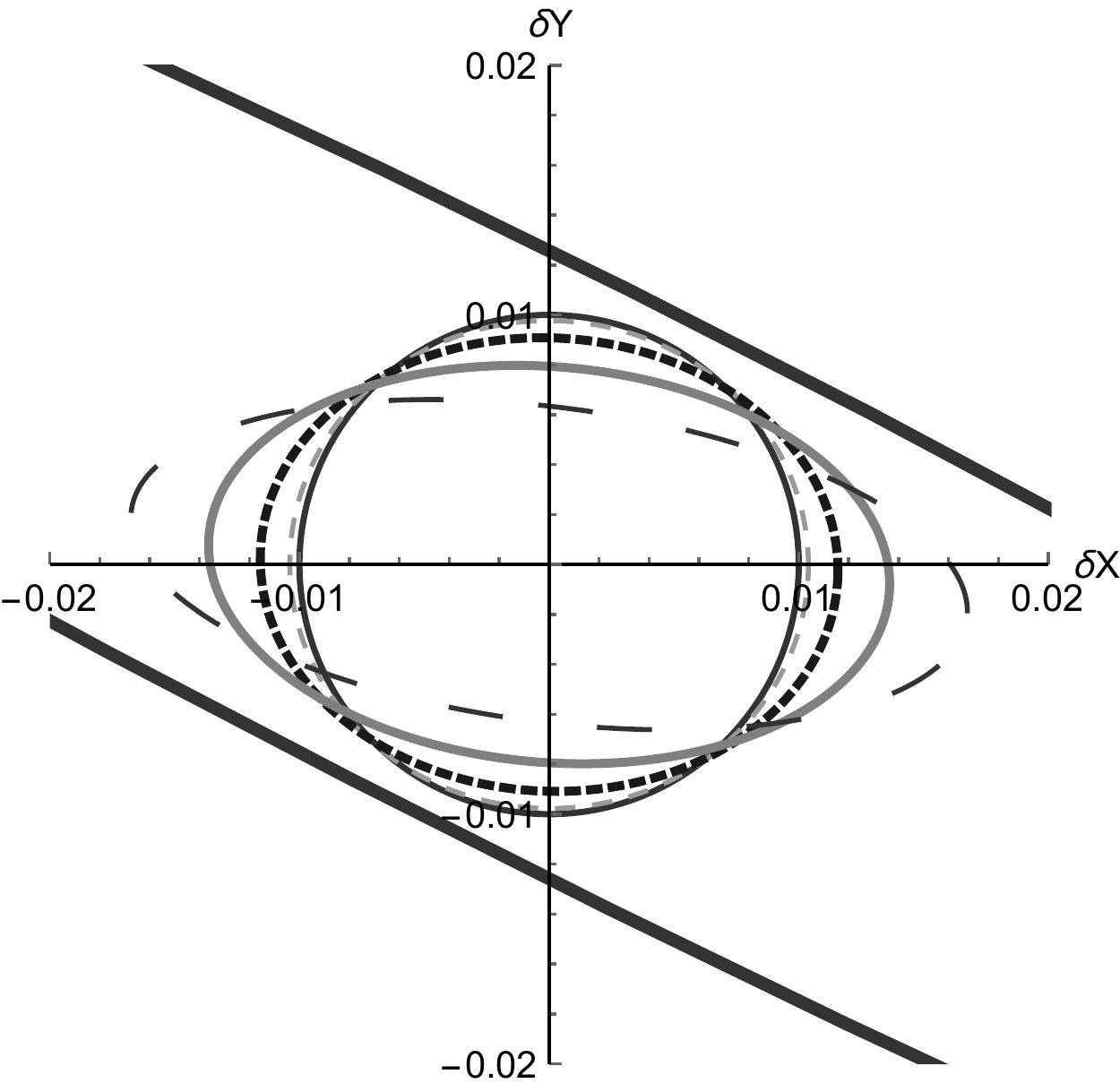}}\\
\subfloat{\includegraphics[width=\textwidth]{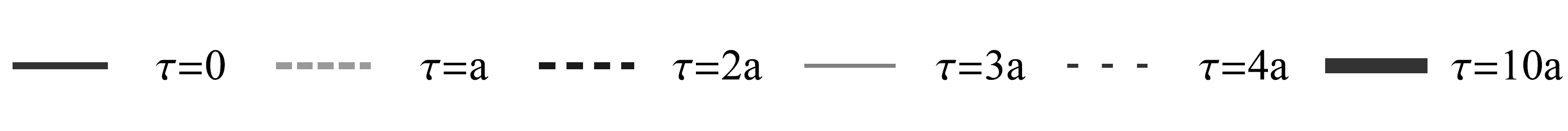}}
\end{center}
\caption{Examples of numerical solutions describing the first-order time evolution of a ring of test particles. Upper left: a stable ring near a static geodesic on the axis, an exact harmonic oscillation; upper right: a converging ring, stable; lower left: a diverging ring in the $\phi$ direction, lower right: a diverging ring in the $\rho$ direction. Axes are scaled by half the distance between the black holes, $a$.}
\label{tab:gt}
\end{figure}
Inspecting \Fref{tab:gt}, we can see that the upper left and upper right orbits are stable while the other two diverge from the original configuration. This is just a hint at the underlying dynamics and one would need to explore a general perturbation of the geodesics. It is, however, obvious that there are both stable and unstable trajectories in the present spacetime. Related to this, we remark that the Hamilton-Jacobi equation is separable for the classical motion in prolate spheroidal coordinates, which however does not translate into the general relativistic situation.
\section{Conclusions}
We have investigated paths of charged test-particles in the background field of two extremally charged black holes held in equilibrium by their electromagnetic field. The spacetime is static and axially symmetric. We note that some of the trajectories we studied do not admit a Newtonian limit since they do not avoid the strong-field regions near the black-hole horizons. It is of interest that as opposed to the Newtonian case, there are regions with two different angular velocities for a single radius of the orbit, which might have observable consequences. This may be the case for all charged, static, asymptotically flat spacetimes as these approximate the Reissner-Nordstr\"{o}m. However, it may happen that the double-frequency region is empty and one would need to study these cases separately. It would then be rather interesting to study the same problem in the Kerr-Newman stationary spacetime. It is to be seen whether the region of double frequencies would be stable under perturbations or not.
\ack
JR was supported by Student Faculty Grant of Faculty of Mathematics and Physics, Charles University in Prague. MZ was supported by The Albert Einstein Center, Project of Excellence No. 14-37086G funded by the Czech Science Foundation.

\end{document}